\documentclass[a4paper,UKenglish,cleveref, autoref, thm-restate]{lipics-v2021}

\nolinenumbers
\hideLIPIcs  
\usepackage[utf8]{inputenc}
\usepackage{mathtools}
\usepackage[colorinlistoftodos]{todonotes}
\renewcommand{\epsilon}{\varepsilon}
\renewcommand{\phi}{\varphi}

\usepackage{algorithm}
\usepackage[noend]{algpseudocode}
\usepackage{amssymb}
\usepackage{amsmath}
\usepackage{amsthm}
\usepackage{pgf,tikz}
\usepackage{pgfplots}
\pgfplotsset{compat=1.16}
\usetikzlibrary{arrows,decorations.pathmorphing,calc,patterns,decorations.pathreplacing,fadings,intersections}

\usepackage{xspace}

\newcommand{\R}{\mathbf{R}}

\title{Online General Knapsack with Reservation Costs}

\author{Elisabet {Burjons}\footnote{corresponding author}}{Serra H\'{u}nter Fellow, Universitat Polit\`{e}cnica de Catalunya, Spain}{elisabet.burjons@upc.edu}{https://orcid.org/0000-0002-1825-0097}{}

\author{Matthias {Gehnen}\footnote{corresponding author}}{RWTH Aachen University, Germany}{gehnen@cs.rwth-aachen.de}{https://orcid.org/0000-0001-9595-2992}{}

\authorrunning{E.~Burjons and M.~Gehnen} 

\Copyright{Elisabet Burjons and Matthias Gehnen} 



\category{} 

\relatedversion{}
\acknowledgements{We want to thank Henri Lotze, Daniel Mock and Peter Rossmanith for initial discussions on this topic.}

\begin{document}

\maketitle

\begin{abstract}
In the online general knapsack problem, an algorithm is presented with an item $x=(s,v)$ of size $s$ and value $v$ and must irrevocably choose to pack such an item into the knapsack or reject it before the next item appears. The goal is to maximize the total value of the packed items without overflowing the knapsack's capacity.

As this classical setting is way too harsh for many real-life applications, we will analyze the online general knapsack problem under the reservation model. Here, instead of accepting or rejecting an item immediately, an algorithm can delay the decision of whether to pack the item by paying a fraction $0\le \alpha$ of the size or the value of the item. This models many practical applications, where, for example, decisions can be delayed for some costs e.g. cancellation fees. We present results for both variants: First, for costs depending on the size of the items and then for costs depending on the value of the items.

If the reservation costs depend on the size of the items, we find a matching upper and lower bound of $2$ for every $\alpha$. On the other hand, if the reservation costs depend on the value of the items, we find that no algorithm is competitive for reservation costs larger than $1/2$ of the item value, and we find upper and lower bounds for the rest of the reservation range $0\le\alpha< 1/2$.
\end{abstract}

\section{Introduction}
Assume you are looking for profitable projects to work on next year. Over time, you will receive offers, each offering some profit, but also needing time for completion. With knowledge of the limited amount of hours you can work within a year, you now need to decide whether to accept an offer or not. Naturally (and sadly), you cannot to wait for every offer, before you need to decide about accepting one. This is the classical setting of an online knapsack problem, where you need to make a selection of items/projects that promise a maximum combined profit but can be completed within the capacity restriction (number of working hours/knapsack size).
While it is known that one cannot hope for any competitive ratio in this setting in theory, in a real-life setting the situation is not hopeless:
For each contract, you might be able to cancel it after accepting - but likely this will not be possible without some contractual penalty or cancellation fee.

The same situation will also occur if you are looking for concerts to visit in the next months: At some point, and usually, only within a limited time frame, you can buy tickets for events you want to attend. Of course, you want to see as many concerts as possible (each weighted by your preference), but you only have a limited budget for buying tickets. Again, you might run into situations, where you need to decide whether to buy a particular ticket or not, without knowing what other options will arise. Again, it is possible to return or sell already purchased tickets, but this also might come with some cancellation or resell fee.

All those problems can be modeled as a general knapsack problem, but the classical variant is often not flexible enough. When adding the option to postpone a decision for some part of its profit, not only it gets more realistic in many real-life settings but, as we will see in this paper, it will also allow us to achieve bounded competitive ratios. In those settings it will make no difference whether we model it with some cancellation fee, or some fee to postpone the decision about accepting, as, as it will turn out, there will not be a single item that would be packed without the risk of being removed later on. Therefore, for each item, either it will be rejected immediately, or reserved for later.
The setting of postponing decisions for some fee was introduced as the reservation model for the online simple knapsack problem~\cite{BBHLR21}, where the value of an item equals its size, and later applied to many other online problems~\cite{BurjonsFGLMR23,BurjonsGLMR21,BurjonsGLMR23}. In such problems, an online algorithm has an option, beyond accepting or rejecting an item, to reserve an item to postpone the decision on it, by paying a cost proportional to the value of the item.

In the \emph{knapsack problem}, given a request sequence $S$, a size
function $s\colon S \to \R$ and a value function $v\colon S \to \R$, an algorithm has the task of finding a subset $K\subseteq S$ such that the sum of sizes of $K$
is less or equal to the so-called \emph{knapsack capacity} (which we assume to be normalized to $1$ in this paper) and the sum of the values of the items, i.e., the total gain, is maximized. The online variant of this problem reveals the items of $S$ piece by piece, with an algorithm having to immediately decide whether to pack a revealed item or discard it. In the online knapsack with reservation, the items are revealed piece by piece as well, however, an online algorithm now has three available options, namely, pack the revealed item, discard the item, or reserve it. After the full request sequence $S$ is revealed an algorithm can decide which reserved items to pack, as long as they fit into the knapsack. Items already packed cannot be discarded at this stage. The reservation costs must then be paid for each reserved item regardless of whether it is used in the final packing or not.

When we apply the reservation model to the general version knapsack problem,
there are two natural ways to do it, either the reservation costs depend on the size or
the value of the items. In this paper, we present upper and lower bounds for both cases, with interesting and diverging results.

Online algorithms are generally analyzed using competitive analysis, introduced by Sleator and Tarjan~\cite{ST84}. In simple terms, one compares the performance of an online algorithm to the performance of an optimal offline algorithm on the same instance. In the case of maximization problems, like the knapsack problem, one can define the strict competitive ratio of an algorithm as follows.

The \emph{strict competitive ratio} of an online algorithm $A$ is the highest ratio of any request sequence $S$ between
the gain of $A$ on $S$ and the gain of an algorithm $\mathtt{OPT}$ solving the problem optimally on the same request $S$,
\[{\rm CR}(\mathtt{A})=\sup_{ S }\left\{\frac{{\rm gain}_{\mathtt{OPT}}(S)}{{\rm gain}_{\mathtt{A}}(S)}\right\}\;.\]

In the reservation model, the reservation costs for an item are a fixed percentage $0\le \alpha$ of the size or the value of the item. Every reserved item has such a cost, regardless of whether the item is ultimately used in the knapsack. Such costs are accounted for in the gain portion of the competitive ratio, where they have to be subtracted from the total value packed into the knapsack by algorithm $\mathtt{A}$.

A first analysis shows that if the reservation is proportional to the item size, the online general knapsack with reservation is strictly not competitive. This is because
an adversary can build instances where the value of every
item is exactly $\alpha$ times its size.
Thus, no algorithm gains anything from reserving an item, and the lower bounds for the general knapsack problem without reservation hold, which means that the problem is strictly not competitive~\cite{Marchetti-SpaccamelaV95}.

The key to this model being not competitive, in the strict sense, stems from the fact that one can adversarially artificially limit the total gain achievable by an algorithm. An extended notion of competitiveness is more natural for this problem, the concept of non-strict competitive ratio.

For maximization problems like knapsack, we formally say that an algorithm is $c$-competitive
if there exists a constant $\beta>0$ such that for every request sequence $S$,
\[{\rm gain}_{\mathtt{OPT}}(S)\le c\cdot {\rm gain}_{\mathtt{A}}(S) +\beta\;.\]

This notion of competitiveness is only reasonable in problems, like the general knapsack, where the achievable gain is unbounded. In problems where the gain is bounded, every algorithm would be $1$-competitive according to such a notion. For instance,
in the online simple knapsack, where the value of an item equals its size, with an additive constant of
$1$ one can claim that every online algorithm is optimal, even without packing anything.

In a setting where the reservation costs are proportional to the item value, we show that algorithms can be competitive also in the strict version.

\subsection{Related Work}The knapsack problem in its offline version is one of the classical optimization problems.
It is known to be NP-hard, even the restricted simple knapsack problem ~\cite{Karp72}. However, it allows fully polynomial approximation scheme algorithms, also for the general case~\cite{IbarraK75}. As the knapsack problem is very classical, many variants are studied, in particular the online version:

When considering the online variant with irrevocable decisions and without any knowledge about the future, no deterministic algorithm can achieve a bounded competitive ratio, even for the simple knapsack~\cite{Marchetti-SpaccamelaV95}. If randomization is allowed, then there exists a simple $2$-competitive algorithm for the simple knapsack using only one randomization bit~\cite{BockenhauerKKR14}, while the general problem still does not allow a bounded competitive ratio~\cite{Yunhong08}.

If one relaxes the blindness of an algorithm, a better competitive ratio can be achieved: One variant is the advice model, where an algorithm can ask an oracle for truthful advice bits~\cite{BockenhauerKKR14}. A recent trend investigates the so-called predictions or machine-learned advice, where the given hints no longer need to be truthful~\cite{im21,boyar22,zeynali21}. Another variant assumes rough item sizes are given in advance, but the actual sizes get revealed in an online manner~\cite{gehnen24}.

Another option is to relax the irrevocability of an algorithm's packing decisions.
The setting, when an algorithm is allowed to discard already packed items is also known as knapsack with removability, which is $\Phi$-competitive for the simple knapsack~\cite{IwamaT02}, where $\Phi$ is the golden ratio. Some models also allow repacking a limited amount of removed items~\cite{boeckenhauer23}. Also, variants when removing items is connected to costs is studied~\cite{HanKM14}, even if only for the simple knapsack.

Furthermore, there is research combining relaxations of blindness and irrevocability. Recently, the knapsack problem with removability was studied in the advice setting~\cite{boeckenhauer24}.

Some variants also relax the packing restriction for an online algorithm, for example, by allowing the algorithm to exceed the knapsack size~\cite{iwama10, han19} or by allowing it to cut items into parts~\cite{han10}.

\subsection{Our Contributions}
In this paper, we present upper and lower bounds for the online general knapsack with reservation costs, when the reservation costs depend both on the size and value of the items.

When the reservation costs depend on the size of the items, we see that, regardless of the reservation costs there is an upper and lower bound of $2$ for the (non-strict) competitive ratio.

\begin{figure}
  \begin{tikzpicture}[scale=0.7]
	\begin{axis}[ ymin=1,xmin=0,xmax=0.5,ymax=500, xlabel=$\alpha$, ylabel=Competitive Ratio, grid=major]
    	\addplot[name path=lb1, domain=0:0.48, dashed] {2*(1-x+sqrt((2-3*x)*x))/(1-2*x)^2};
    	\addplot[name path=ub3, domain=0:0.48] {2*(2*x+sqrt(x*(1+2*x))*(1+2*x+2*sqrt(x*(1+2*x)))/((1-2*x)^2*sqrt(x* (1+2*x)))};
	\end{axis}
  \end{tikzpicture}
\hspace*{2em}
   \begin{tikzpicture}[scale=0.7]
	\begin{axis}[ ymin=1,xmin=0,xmax=0.1,ymax=7, xlabel=$\alpha$, ylabel=Competitive Ratio, grid=major]
    	\addplot[name path=lb1, domain=0:0.1, dashed] {2*(1-x+sqrt((2-3*x)*x))/(1-2*x)^2};
    	\addplot[name path=ub3, domain=0.001:0.1] {2*(2*x+sqrt(x*(1+2*x))*(1+2*x+2*sqrt(x*(1+2*x)))/((1-2*x)^2*sqrt(x* (1+2*x)))};
	\end{axis}
  \end{tikzpicture}

\caption{A schematic plot of the upper and lower bounds for reservation depending on item value. The dashed line represents the general lower bound proven in Theorem~\ref{thm:vlb} and the continuous line is the upper bound achieved in Corollary~\ref{thm:vub-mark} by analyzing Algorithm~\ref{alg:submult}.}
\label{fig:plot}
\end{figure}
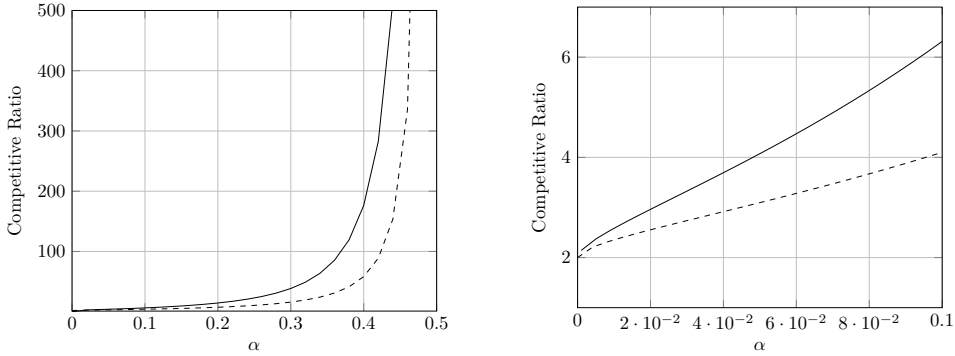

For the reservation costs depending on the value of the items, we get the upper and lower bounds depicted in Figure~\ref{fig:plot}. The upper bound is
 \[\frac{2 (2 \alpha + \sqrt{\alpha (1 + 2 \alpha)})(1 + 2 \alpha + 2 \sqrt{\alpha (1 + 2 \alpha)})}{(1 - 2 \alpha)^2 \sqrt{\alpha (1 + 2 \alpha)}}\;,\] whereas the lower bound is
 $\frac{2(1+\sqrt{\alpha(2-3\alpha)}-\alpha)}{(1-2\alpha)^2}$.
 
We can see that the competitive ratio is $2$ when the reservation costs go to $0$ and the problem becomes not competitive for reservation costs of $0.5$ times the value of the item. In between those scenarios we have upper and lower bounds that do not match but that are asymptotically similar.

\section{Reservation costs proportional to item size}

As we already mentioned in the introduction, when the reservation costs depend on the size of an item, every online algorithm is strictly not competitive.

Let us expand on this idea. By presenting items of the type $x=(s,\alpha s)$ for any size $s$ between $0$ and $1$, an adversary makes it so that every reservation results in a net gain of $0$ if the item gets reserved and then used, and in a negative gain if the item is not used. Thus, no reasonable algorithm reserves any item.

However, any items that directly get taken into the knapsack make the algorithm strictly not competitive. This is because the lower bound approach for the online general knapsack~\cite{Marchetti-SpaccamelaV95} can be applied here. Namely, if the first offered item is not taken, no more items will appear, rendering such algorithms uncompetitive. However, if the first offered item is placed into the knapsack, an item $x=(1,V)$ will be offered of size $1$ and arbitrary value $V$, which would be taken by the optimal solution but cannot be taken by an algorithm with a partially occupied knapsack, rendering all such algorithms uncompetitive as well.

Thus, in this section, we prove that the general knapsack problem with reservation has a non-strict competitive ratio of $2$ by providing matching upper and lower bounds in all of the range of reservation costs, namely for any $\alpha>0$.

\subsection{Upper Bound}
Given an item $x=(s_x,v_x)$ that has size $s_x$ and value $v_x$, we talk about the density of an item as the ratio between its value and size, namely $d(x)=v_x/s_x$, and for a given set of items $X$ we define its value $v(X)=\sum_{x\in X} v_x$ and size $s(X)=\sum_{x\in X} s_x$ and accordingly its density as the ratio between them $d(X)=v(X)/s(X)$. To build an algorithm for the upper bound we will consider at all times the densest knapsack $R_s$, i.e., from the reserved item set take the subset containing the densest items that add up to 1 or more. Observe that $R_s$ might not be packable as is, namely, the least dense item in $R_s$, might not fully fit inside the knapsack if we wanted to pack the densest reserved items.
We also define $x^{R_s}_{\delta}$ or to abbreviate $x_{\delta}$ to be the least dense item in $R_s$, and $d^{R_s}_{\delta}$ or $d_{\delta}$ its density.

Let us consider Algorithm~\ref{alg:sub} for online general knapsack. This algorithm reserves items as long as they are ``dense enough''. This means, that if an item has a value less than $\alpha$ times its size it is not reserved, as shown in line~\ref{line:lowdensity}. If, on the other hand, the knapsack is not full yet, and the item is denser than $\alpha$, the item is reserved, as shown in line~\ref{line:reservesmall-s}. Then, in line~\ref{line:reservegeneral}, the algorithm reserves items as long as they are $c$ times more dense than $d_{\delta}$, where $c$ is a chosen constant $c=1+\epsilon$, with $\epsilon>0$.
\begin{algorithm}[tb]
  \begin{algorithmic}[1]

  \State{$R_s\coloneqq\emptyset$;}\Comment{Densest reserved items up to total size 1}
  \State{$d_{\delta}^{R_s}\coloneqq0$; }\Comment{Smallest density of all items in $R_s$}
  \For{$k=1,\ldots,n$}
	\If{$d(x_k)\le \alpha$} {reject $x_k$;}\label{line:lowdensity}
  	\ElsIf{$s(R_s)<1$} {reserve $x_k$;}\label{line:reservesmall-s}
		\State{$R_s\coloneqq R_s\cup x_k$;}
		\Repeat
		 \State{Define $x_{\delta}$ as least dense item in $R_s$;}
		 \If{$s(R_s\setminus x_{\delta})\ge1$} {$R_s\coloneqq R_s\setminus x_{\delta}$;}
		 \EndIf
		\Until{$s(R_s\setminus x_{\delta})<1$;}
		\State{Set $d_{\delta}^{R_s}$ as density of least dense item in $R_s$;}
  	\ElsIf{$d(x_k)\ge c\cdot d_{\delta}^{R_s}$} {reserve $x_k$;}\label{line:reservegeneral}
		\State{$R_s\coloneqq R_s\cup x_k$;}
		\Repeat
		 \State{Define $x_{\delta}$ as least dense item in $R_s$;}
		 \If{$s(R_s\setminus x_{\delta})\ge1$} {$R_s\coloneqq R_s\setminus x_{\delta}$;}
		 \EndIf
		\Until{$s(R_s\setminus x_{\delta})<1$;}
		\State{Set $d_{\delta}^{R_s}$ as density of least dense item in $R_s$;}
  	\Else{ reject $x_k$;}\label{line:reject-s}
	\EndIf
  \EndFor
  \State{pack the reserved items optimally}\label{line:packatend-s}
  \end{algorithmic}
  \caption{Algorithm for an upper bound when the reservation costs depend on the size of an item. Here, $c$ is defined as $c=1+\delta$ for a small $\delta>0$.  $R_s$ always contains the densest items of total size 1 or larger.}
  \label{alg:sub}
\end{algorithm}
\newpage
\begin{theorem}\label{thm:sub}
 
 For every reservation factor $\alpha$ and $\varepsilon>0$ the factor $c$ can be chosen such that the competitive ratio of Algorithm~\ref{alg:sub} is smaller than $2+\varepsilon$.
\end{theorem}

\begin{proof}
 If no item in a request sequence has a density $d(x_k)>\alpha$ this means, necessarily that every possible combination of items filling the knapsack has a value less than $\alpha\cdot 1$, i.e., the density times the size of the knapsack. The competitive ratio of Algorithm~\ref{alg:sub} in such instances is $1$, if we take $\beta>\alpha$ in the competitive ratio. That is, if $S$ is one such request sequence
 \[{\rm gain}_{\mathtt{OPT}}(S)\le \alpha \le  {\rm gain}_{\mathtt{A}}(S) +\beta\;,\]
 for any $\beta\ge\alpha$ regardless of the actual gain of the algorithm, which in this case is $0$, as such items are rejected in line~\ref{line:lowdensity}.
 
 If there are some items with a density larger than $\alpha$, but they do not fill the whole knapsack, these will be reserved in line~\ref{line:reservesmall-s}. If no item triggers lines~\ref{line:reservegeneral} or~\ref{line:reject-s}, the knapsack $R_s$ might have a total size of more than $1$ but only the last item exceeded the capacity. If $s(R_s)\le 1$, then Algorithm~\ref{alg:sub} can pack all of $R_s$ into the knapsack and the optimal solution may additionally have items of density $\alpha$ up to the full knapsack size. Thus,
 \[{\rm gain}_{\mathtt{OPT}}(S)\le v(R_s)+\alpha\le {\rm gain}_{\mathtt{A}}(S) -\alpha +\beta\;,\]
 for any $\beta\ge 2\alpha$. If, on the other hand 
 $s(R_s)>1$, ${\rm gain}_{\mathtt{OPT}}(S)\le v(R_s)$ and, because only the last item might exceed the capacity, Algorithm~\ref{alg:sub} might not be able to fill the full knapsack but certainly will be able to get ${\rm gain}_{\mathtt{A}}(S)\ge \frac12 v(R_s)$. This is because all the densest items except for the least dense one, $x_{\delta}$, fit into the knapsack together, and also $x_{\delta}$ alone must fit the knapsack, and, either $v(R_s)-v(x_{\delta})\ge\frac12 v(R_s)$ or $v(x_{\delta})\ge \frac12 v(R_s)$.
 \[{\rm gain}_{\mathtt{OPT}}(S)\le v(R_s) = 2\cdot\frac{v(R_s)}{2}= 2\left(\frac{v(R_s)}{2}-\alpha s(R_s)\right) + 2\alpha s(R_s) <2\cdot {\rm gain}_{\mathtt{A}}(S) +4\alpha\;,\]
 as $s(R_s)<2$ by construction. Thus, in this case, the competitive ratio is $2$ with any $\beta \ge 4\alpha$.
 
 If line~\ref{line:reservegeneral} is processed during the execution of Algorithm~\ref{alg:sub}, each time the algorithm has reserved an extra total size of at least $1$ and at most $2$, the minimum density of $R_s$ increases by at least a factor of $c$, thus, after processing the full input, if $c^{k-1}\alpha < d_{\delta}\le c^k\alpha$ the reservation costs must be at most $\alpha\cdot 2(k+1)$, i.e., at most $2$ knapsacks for each reservation density. The algorithm packs at least $\frac 12 v(R_s)$, as in the previous case. We also consider $v_c$ the total value of items in $R_s$ which are denser than $c\cdot d_{\delta}$, and $s_c$ their total size, and we see that the algorithm packs, also, at least $v_c$, this is because the items of density at least $c\cdot d_{\delta}$ do not overflow $R_s$ and thus, can be packed together into the knapsack. With these two values, we can also state that $\frac 12 v(R_s)\ge \frac{v_c + (1 - s_c)d_{\delta}}{2}$, because the total value $v(R_s)$ must contain a full knapsack with value $v_c$ for the densest items and the topmost item having density $d_{\delta}$, which could occupy at most the rest of the knapsack.
 Thus, we can write
 \[{\rm gain}_{\mathtt{A}}(S)\ge \max\left\{\frac{v_c + (1 - s_c)d_{\delta}}{2}, v_c\right\}-2\alpha(k+1)\;.\]
 
 The gain of the optimal solution is made up of reserved and not reserved items, the non-reserved items have density at most $(c-\epsilon)d_{\delta}$, and the reserved items must be part of $R_s$ and denser than $c\cdot d_{\delta}$ to be optimal, thus in total we have
 \[{\rm gain}_{\mathtt{OPT}}(S)\le v_c + (c-\epsilon)d_{\delta}(1-s_c)<v_c +c\cdot (1-s_c)d_{\delta}\;.\]
 
 If $\max\left\{\frac{v_c + (1 - s_c)d_{\delta}}{2}, v_c\right\}=v_c$, we can compute,
 \begin{align*}
   2 \cdot {\rm gain}_{\mathtt{A}}(S) +\beta &\ge
   2(v_c - 2\alpha(k+1))+ \beta\\
  &\ge v_c + c\cdot (1 - s_c)d_{\delta} \left(\frac{v_c}{c\cdot (1 - s_c)d_{\delta}} - 4\alpha\frac{k}{c\cdot (1 - s_c)d_{\delta}}\right)-4\alpha+\beta
 \end{align*}
 Take into account that $v_c\ge\frac{v_c + (1 - s_c)d_{\delta}}{2}$ by construction, thus, $v_c\ge (1-s_c)d_{\delta}$, and $(1-s_c)d_\delta>d_{\delta}>c^{k-1}\alpha$. Now we can set $\beta\ge 4\alpha+\beta'$ and we obtain
 \[2 \cdot {\rm gain}_{\mathtt{A}}(S) +\beta \ge v_c + c(1 - s_c)d_{\delta}\left(\frac{1}{c} - 4\frac{k}{c^k}\right)+\beta'\]
 
 On the other hand, if $\max\left\{\frac{v_c + (1 - s_c)d_{\delta}}{2}, v_c\right\}=\frac{v_c + (1 - s_c)d_{\delta}}{2}$, we can compute,
 \begin{align*}
   2 \cdot {\rm gain}_{\mathtt{A}}(S) +\beta &\ge
   2\left(\frac{v_c + (1 - s_c)d_{\delta}}{2} - 2\alpha(k+1)\right)+ \beta\\
  &\ge (v_c + c(1 - s_c)d_{\delta}) \left(\frac{v_c + (1 - s_c)d_{\delta}}{v_c + c(1 - s_c)d_{\delta}} - 4\alpha\frac{k}{v_c + c(1 - s_c)d_{\delta}}\right)-4\alpha+\beta\\
  &\ge (v_c + c(1 - s_c)d_{\delta}) \left(\frac{1}{c} - 4\alpha\frac{k}{v_c + c(1 - s_c)d_{\delta}}\right)-4\alpha+\beta
 \end{align*}
 Take into account that $v_c\le\frac{v_c + (1 - s_c)d_{\delta}}{2}$ by construction, thus, $v_c\le (1-s_c)d_{\delta}$, and $(1-s_c)d_\delta>d_{\delta}>c^{k-1}\alpha$. Now we can set $\beta\ge 4\alpha+\beta'$ and we obtain
 \[2 \cdot {\rm gain}_{\mathtt{A}}(S) +\beta \ge (v_c + c(1 - s_c)d_{\delta})\left(\frac{1}{c} - 4\frac{k}{c^{k-1}+c^k}\right)+\beta'\]
 
 We can now see that for every $\epsilon>0$ there exists a $K_0$ and a value of $c=1+\delta$ for some $\delta>0$ such that for every $k>K_0$,
 \[(v_c + c(1 - s_c)d_{\delta})\left(\frac{1}{c} - 4\frac{k}{c^{k-1}+c^k}\right)+\beta'\ge v_c + c(1 - s_c)d_{\delta}-\epsilon+\beta'\]
 and also
 \[v_c + c(1 - s_c)d_{\delta}\left(\frac{1}{c} - 4\frac{k}{c^k}\right)+\beta'\ge v_c + c(1 - s_c)d_{\delta}-\epsilon+\beta'\]
 for those values of $k>K_0$, one can set $\beta'>\epsilon$ and bound
 \[2 \cdot {\rm gain}_{\mathtt{A}}(S) +\beta \ge v_c + c(1 - s_c)d_{\delta}-\epsilon+\beta'\ge {\rm gain}_{\mathtt{OPT}}(S)\]
 
 For every $k\le K_0$, because $K_0$ is a constant, we can go back a few steps and see that
 \begin{align*}
   2 \cdot {\rm gain}_{\mathtt{A}}(S) +\beta &\ge
   2\left(\max\left\{\frac{v_c + (1 - s_c)d_{\delta}}{2}, v_c\right\} - 2\alpha(k+1)\right)+ \beta\\
  &\ge (v_c + c(1 - s_c)d_{\delta})\frac{1}{c} - 4\alpha k+\beta'
 \end{align*}
 so if we set $\beta'>4\alpha K_0+\epsilon$ the bound also fits in this case and we have proved the desired competitive ratio.
\end{proof}

\subsection{Lower bound}
We can achieve a matching lower bound of $2$ by constructing an appropriate adversarial instance as follows.

\begin{theorem}
Given a reservation cost of $0<\alpha\le 1$ for the item size, no online algorithm 
solving the online knapsack problem can be 
$c$-competitive for any constant $c< 2$.
\end{theorem}

\begin{proof}
Given a reservation cost $\alpha$, let us consider
the following adversarial request sequence.

The first item is $x_1=(\frac12+\epsilon,C)$ for some $\epsilon>0$ and some large value of $C$.
If the item is taken, an item $x_2=(1,3C)$ is presented and the request sequence ends.
Otherwise, if the item is rejected no other item arrives and the request sequence ends.
However, the item can also be reserved, if this is the case, items 
$x_i=(\frac12 +\epsilon^i, C)$ are presented 
until either $x_i$ is taken for some value of $i$ and an item $(1,3C)$ is presented,
or $x_i$ is rejected for some $i$, and a complementary item $x_i'=(\frac12-\epsilon^i, C)$ arrives. The adversary will present items $x_i$ until one is taken or rejected or until the sum of the total reservation costs adds up to $C$.

If the request sequence terminates after an item $(1,3C)$, the algorithm achieves a gain of at most $C$,
but the optimal gain is at most $3C$, this means that for any constant $\beta>0$, we can choose $C$ such that the competitive ratio is larger than $2$.

If the request sequence terminates after an item $x_i$ is rejected, the optimal takes $x_i$ and $x_i'$ into the knapsack, with an added value of $2C$, where the algorithm can only pack a total value of $C$, achieving again a competitive ratio as close to $2$ as necessary for an arbitrary choice of $C$ that overcomes the value of any constant $\beta$.

If the request sequence terminates because the first item is rejected, we can choose $C>2\beta$ and get a competitive ratio larger than $2$.

Otherwise, if the algorithm keeps reserving every offered item, at some point the reservation costs become than $C$, at which point the request sequence ends, the gain of such an algorithm is negative and the desired competitive ratio is also achieved by choosing a $C>2\beta$.
\end{proof}

\section{Reservation costs proportional to item value}
In this section, we present upper and lower bounds that depend on the value of the item. First, we see that, in this case, one can translate some bounds from the online simple knapsack. As announced, we use the strict version of the competitive ratio in this setting.

\begin{observation}\label{obs:simpleknapsack}
 The online general knapsack with reservation costs depending on the value is reducible to the online simple knapsack with reservation costs. 
\end{observation}

 If the reservation costs depend on the value, an adversary can present an instance with fixed density $1$, i.e., the items are $x=(s,s)$ and mimic a simple knapsack instance.
 Thus, no algorithm for online general knapsack with reservation costs depending on the value can achieve a better competitive ratio than the bounds for the online simple knapsack with reservation costs.

This observation is valid for both the strict and non-strict versions of the competitive ratio because if instead of choosing density $1$ we choose an arbitrary density $d$, one can offset any additive constant $\beta$ when computing the competitive ratio. For this reason, all of the bounds in this section will be calculated as strict ratios, as one can always scale the density of all of the offered items to overcome $\beta$.

We cannot make a similar claim to Observation~\ref{obs:simpleknapsack} for the upper bound, as no algorithm is guaranteed to receive an instance where the density is fixed.

The lower bounds offered by applying Observation~\ref{obs:simpleknapsack}, however, are not as good as what can be achieved by considering instances where the items have different densities.

\subsection{Lower Bound}
To build lower bounds, we build a set of adversarial instances such that no algorithm can achieve a better ratio than the proposed bound. A few observations will help to analyze the behavior of these algorithms.

\begin{observation}\label{obs:take}
 No algorithm packing items before the end of the instance is competitive.
\end{observation}
 
If an algorithm decides to pack an item before the end of the request sequence has been announced, the adversary can always craft an item $x$ of size $1$ and arbitrary value which will render such an algorithm uncompetitive. Because the value of $x$ is arbitrary, this holds true for both in the strict and non-strict versions of the competitive ratio.

Thus, we do not need to consider algorithms that preemptively pack during our lower bound analysis. Similarly, we do not have to consider algorithms that do not reserve the first item.

\begin{observation}\label{obs:nores}
 Any algorithm not reserving any item is uncompetitive.
\end{observation}

If an algorithm decides not to reserve the first offered item, the adversary can always stop the reservation sequence there, which will render such an algorithm strictly uncompetitive. For the non-strict version, an adversary can offer increasingly denser items of the same size until an arbitrary value is achieved. If an algorithm does not reserve any of these items the algorithm cannot be competitive for any fixed value of $\beta$ in the non-strict sense.
Thus, by scaling the density of the first reserved item, we can assume that the first item offered by the adversary is reserved.

With these considerations, let us now present the general bound for all other algorithms.

\begin{theorem}\label{thm:vlb}
	No algorithm for the online general knapsack can achieve a competitive ratio smaller than $\frac{2(1-\alpha+\sqrt{(2-3\alpha)\alpha})}{(1-2\alpha)^2}$ for a reservation factor of $0<\alpha<0.5$.
\end{theorem}
\begin{proof}
	Let $\epsilon_1, \epsilon_2 >0$ and $N$ a natural number.
	The adversarial instances consist of items $x_k\coloneqq (1-k \epsilon_1, v_k)$ for $k\geq 0$, starting with $x_0\coloneqq (1,1)$. Each value $v_k$ is larger than its predecessor by a decreasing factor $f_k$, so $v_k \coloneqq f_1 f_2 \dots f_k$.
	Here $f_1$ is chosen sufficiently large (e.g., by a factor of $(1+\epsilon_2)$ larger than the aimed competitive ratio $\frac{2(1-\alpha+\sqrt{(2-3\alpha)\alpha})}{(1-2\alpha)^2}$). Furthermore, $f_k$ decreases slowly towards a factor of $1$, such that at any point $k>N$ the factor $f_k$ is at least $(1-\epsilon_2) f_{k-N}$. If an algorithm decides to not reserve an item $x_k$, the adversary stops presenting further items of this form.
    
	If the decision to not reserve the item $x_k$ was made for $k=0$, the instance ends and we have the situation described in Observation~\ref{obs:nores}.
    
	If the algorithm decides to not reserve an item $x_k$ for $0<k<N$, the instance ends as well. The optimal solution can consist of $x_k$, which, by construction is at least $\frac{2(1-\alpha+\sqrt{(2-3\alpha)\alpha})}{(1-2\alpha)^2}$ times larger than the largest reserved item from the algorithm $x_{k-1}$. Therefore, the competitive ratio in this case also exceeds the claimed one.
    
	If an algorithm decides to not reserve an item $x_k$ for $k\geq N$, the algorithm presents an item $y\coloneqq (k\epsilon_1, v_{k-1})$. If the algorithm decides to reserve $y$, the instance stops; otherwise, copies of $y$ will be presented as long as either a copy gets reserved or more than $\frac{2(1-\alpha+\sqrt{(2-3\alpha)\alpha})}{(1-2\alpha)^2}$ copies of $y$ got presented.
	If no copies of $y$ got reserved, the optimal solution can pack all of them together, resulting in a packing worth $\frac{2(1-\alpha+\sqrt{(2-3\alpha)\alpha})}{(1-2\alpha)^2} v_{k-1}$ while the algorithm only achieves $v_{k-1}$. Even ignoring the reservation costs the algorithm has paid so far, the algorithm can not achieve the desired competitive ratio. Note that this assumes $\epsilon_1$ to be sufficiently small, which can be achieved by choosing $\epsilon_1$ accordingly for fixed $N$ and $\epsilon_2$.
	Therefore, we assume that the algorithm will reserve one copy of $y$.
    
	In this case, the optimal solution consists of the item $x_k$ together with $y$, achieving $v_k+v_{k-1}$. The algorithm can pack either $y$ or $x_{k-1}$, both with value $v_{k-1}$. Furthermore, the algorithm will have to pay for reservation costs of $\alpha v_{k-1}$ for $y$ plus
	\[\alpha \sum_{i=0}^{k-1} v_i \geq \alpha \sum_{i=k-N}^{k-1} v_i \geq \alpha v_{k-N} \sum_{i=0}^{N-1} f_k^i = \alpha v_{k-N} \frac{1-f_k^{N}}{1-f_k}\]
 	for the reserved items $x_i$.
	 
	Therefore, the algorithm achieves a competitive ratio of
	\[ \frac{v_k+v_{k-1}}{v_{k-1} -\alpha v_{k-1} - \alpha v_{k-N} \frac{1-f_k^{N}}{1-f_k}} =
 	\frac{f_k+1}{1 -\alpha - \alpha \frac{v_{k-N}}{v_{k-1}} \frac{1-f_k^{N}}{1-f_k}}\]
    
	For every $\epsilon_3$, the geometric sum can be bounded as a series for a sufficiently large chosen $N$, resulting in a competitive ratio of
	\[ \frac{f_k+1}{1 -\alpha - \alpha (\frac{1}{1-1/f_k}-\epsilon_3)}\;.\]
    
	A factor $f_k = \frac{-1+\alpha - \sqrt{2 \alpha- 3 \alpha^2}}{-1+2\alpha}$ promises the minimal competitive ratio among all possible factors. The resulting competitive ratio for this $x_k$ turns out to be exactly
	\[\frac{2(1-\alpha+\sqrt{(2-3\alpha)\alpha})}{(1-2\alpha)^2}\;.\] Note that this minimum exists as it is defined by a trade-off between high total reservation costs (in case of a smaller factor $f$) and of course the ratio between the algorithms item and the large item the optimal solution can use (in case of larger factor $f$).
    
	Finally, note that if the algorithm does reserve all items $x_k$, at some point it will reserve some item $x_y$ with $f_y < 1+\epsilon_2$. Then the instance ends. The algorithm as well as an optimal solution can pack at most the value $v_k$ of the last presented item, however, from the algorithm's value we need to deduct the reservation costs of at least $\alpha N (v_k - 2\epsilon)$ for the last $N$ items. With sufficiently large $N$, the algorithm does not achieve a bounded ratio.
\end{proof}

\subsection{Upper Bounds}

In this section we see that Algorithm~\ref{alg:submult}, when analyzed with reservation costs depending on the item value, provides an upper bound that goes to $2$ for small values of $\alpha$, and diverges at $\alpha=\frac12$ at the same asymptotical rate as the lower bound.

\begin{algorithm}[tb]
	\begin{algorithmic}[1]

		\State{$R_s\coloneqq\emptyset$;}\Comment{Densest reserved items up to total size 1}
		\State{$d_{\delta}^{R_s}\coloneqq0$; }\Comment{Smallest density of all items in $R_s$}
		\For{$k=1,\ldots,n$}
		\If{$s(R_s)<1$} {reserve $x_k$;} 
		\State{$R_s\coloneqq R_s\cup x_k$;}
		\Repeat
		\State{Define $x_{\delta}$ as least dense item in $R_s$;}
		\If{$s(R_s\setminus x_{\delta})\ge1$} {$R_s\coloneqq R_s\setminus x_{\delta}$;}
		\EndIf
		\Until{$s(R_s\setminus x_{\delta})<1$;}
		\State{Set $d_{\delta}^{R_s}$ as density of least dense item in $R_s$;}
		\ElsIf{$d(x_k)\ge c\cdot d_{\delta}^{R_s}$} {reserve $x_k$;} 
		\State{$R_s\coloneqq R_s\cup x_k$;}
		\Repeat
		\State{Define $x_{\delta}$ as least dense item in $R_s$;}
		\If{$s(R_s\setminus x_{\delta})\ge1$} {$R_s\coloneqq R_s\setminus x_{\delta}$;}
		\EndIf
		\Until{$s(R_s\setminus x_{\delta})<1$;}
		\State{Set $d_{\delta}^{R_s}$ as density of least dense item in $R_s$;}
		\Else{ reject $x_k$;} 
		\EndIf
		\EndFor
		\State{pack the reserved items optimally}\label{line:packatend-smult}
	\end{algorithmic}
	\caption{Algorithm for an upper bound in the setting of reservation costs depending on the value for a given $c\geq 1$.  $R_s$ always contains the densest items of total size 1 or larger.}
	\label{alg:submult}
\end{algorithm}
Note that Algorithm~\ref{alg:submult} is almost identical to Algorithm~\ref{alg:sub}, and only differs by a different factor $c$ and for the handling of the least dense items. Algorithm~\ref{alg:submult} reserves items that are less dense than $\alpha$, which Algorithm~\ref{alg:sub} does not do (see line \ref{line:lowdensity}). Thus, Algorithm~\ref{alg:submult} achieves the claimed competitive ratio in a strict sense. 
\newpage
\begin{theorem}\label{thm:ub-mark}
Algorithm~\ref{alg:submult} achieves a strict competitive ratio of at least
  \[ \frac{2c}{1 - \alpha(\frac{4}{1-\frac{1}{c}}-2)} \;.\] for a given $c>1$ and $0 < \alpha < \frac{1}{2}$.
\end{theorem}
\begin{proof}
 The analysis is similar to the one in \cref{thm:sub}, with the appropriate change in the reservation costs. In this case, however, we can calculate the strict competitive ratio, which makes the final calculations more straightforward, as we will see.

 Without loss of generality, assume that $d_{\delta}=1$ at the end of the algorithm. All of the items with a density larger than $c$ have been reserved and fit in $R_v$. Let us say all items with a density larger than $c$ have a combined value of $v$ and a combined size of $s<1$.
 There can be items with density in the interval $(1,c]$ that are also in $R_v$ and have a combined value of $w$ and a size of $t$, with $t+s<1$. 
	 
 Algorithm~\ref{alg:submult}, just like Algorithm~\ref{alg:sub}, is able to pack half of the value of $R_v$, as we argued in \cref{thm:sub}, resulting in a packing with value at least $\frac{v+w+(1-s-t)}{2}$, but we can also be sure that the solution is at least $v+w$ by only taking the items of density larger than $1$. Additionally, if the algorithm reserves an amount of $x$ of items with density $1$, the algorithm is able to pack a value of at least $\frac{x}{2}$.
 
 The reservation costs are the sum of the reservation costs of $v$ and $w$ and the reservation costs of the less dense items.
 The costs for the dense items are simply $\alpha (v+w)$. For items of density exactly $1$ we can have at most value $x\leq 2-t$, which takes into account the overflowing of $R_v$ at a time when the items that make up $v$ might not have arrived. The reservation costs for the smaller items can be calculated by a geometric sum; and, as at most a size of $2$ can be reserved until the marker was increased by a factor of $c$, are at most $\alpha 2\sum_{i=1}^k \frac{1}{c^i}\le \alpha(\frac{2}{1-\frac{1}{c}}-2)$. 
    
 The optimal solution can consist of reserved and unreserved items. With unreserved items having density at most $c$, the optimal solution can not be larger than $v+(1-s)c$.
    
 Together, the algorithm achieves a competitive ratio of at least
 \[\frac{v+(1-s)c}{\max\left\{\frac{v+w+1-s-t}{2}, v+w,\frac{x}{2}\right\} - \alpha(v+w+x+ \frac{2}{1-\frac{1}{c}}-2) }\;.\]
	 
 For an upper bound analysis, and given a factor $c$ and $\alpha$, we need to minimize this ratio. On the other hand, $v$, $w$, $s$, and $t$ are chosen adversarially to maximize the ratio.
 
 First, note that the ratio is maximized when $w=t=0$, as $w$ only increases the gain of the algorithm and a higher $t$ only contributes even negatively to the reservation costs due to the restriction of $x\leq 2-t$. For this, observe that the algorithm packs at least $v+w$ into the knapsack, which means that by subtracting the reservation costs, the $w$ contribution is $(1-\alpha)w$, which is always positive.
 Therefore, the ratio looks like this:
 \[\frac{v+(1-s)c}{\max\left\{\frac{v+1-s}{2}, v, \frac{x}{2}\right\} - \alpha(v+x+\frac{2}{1-\frac{1}{c}}-2)}\;,\]
 where we integrate the $2\alpha$ term into the geometric sum.
 
 Observe that if $\max\left\{\frac{v+1-s}{2}, v\right\}=v$, $v\ge 1-s$ whereas $\max\left\{\frac{v+1-s}{2}, v\right\}=\frac{v+1-s}{2}$ when $v\le 1-s$.
 If we take the derivative with respect to $v$ of both of these ratios we see that it is always negative for any $c>1$ and $0\le\alpha\le\frac12$. Thus, to maximize the ratio, every adversarial instance will not present items denser than $c$.
 One can also see this in the following way. Even when comparing just the packing of the algorithm with the optimal solution, one easily sees that having dense items favors the algorithm: Not only is the ratio between $v$ and $v/2-\alpha v$ smaller than the ratio between $(1-s)c$ and $(1-s)/2 - \alpha(1-s)$, which alone would be a reason to set $v$ (and therefore $s$) to zero when trying to achieve a worst possible ratio, it also gets more supported by the fact that the algorithm still has to pay further reservation costs (which therefore lessens the ratio even more, when we cut out profit for both the algorithm and the optimal solution). Therefore, the ratio is largest when no items denser than $c$ appear in the request sequence (namely $v=0$ and $s=0$):
 \[\frac{c}{\max\left\{\frac{1}{2}, \frac{x}{2}\right\} - \alpha(x + \frac{2}{1-\frac{1}{c}}-2)}\;.\]
 At this point, it is easy to see, that the worst case for an algorithm and its competitive ratio is an instance that consists only of items with a total size of $1$ with density $1$ (for example by two items of size $\frac{1}{2}+\varepsilon$ each). This is, as the gain of an algorithm increases when allowing him to pack more items of density $1$, even if twice the reservation costs are deducted (as we only consider $\alpha < 0.5$ here). Note that $x$ must be at least one as $1$ is the smallest density in $R_s$ at the end of the request sequence, thus if $s=w=0$ it must mean $R_s$ only contains items of density $1$.
 
 Therefore, given a fixed $c$, the competitive ratio is 
  \[\frac{c}{\frac{1}{2} - \alpha(1 + \frac{2}{1-\frac{1}{c}}-2)} = \frac{2c}{1 - \alpha(\frac{4}{1-\frac{1}{c}}-2)} \;.\]

\end{proof}
\newpage

\begin{corollary}\label{thm:vub-mark}
	Algorithm~\ref{alg:submult} achieves a strict competitive ratio of at least
	\[\frac{2 (2 \alpha + \sqrt{\alpha (1 + 2 \alpha)})(1 + 2 \alpha + 2 \sqrt{\alpha (1 + 2 \alpha)})}{(1 - 2 \alpha)^2 \sqrt{\alpha (1 + 2 \alpha)}}\;.\] for all $0 < \alpha < \frac{1}{2}$ when $c=\frac{2 \sqrt{2\alpha^{2} + \alpha} + 2\alpha + 1}{1-2\alpha}$.
\end{corollary}
\begin{proof}
	By Theorem~\ref{thm:ub-mark}, we know that Algorithm~\ref{alg:submult} achieves a competitive ratio of
	  \[\frac{2c}{1 - \alpha(\frac{4}{1-\frac{1}{c}}-2)} \;.\]
	for a given $c >1$.
	
	Using calculus, we can fix an optimal version of Algorithm~\ref{alg:submult} by choosing the best factor $c$ for each reservation cost $\alpha$. We take the derivative of the competitive ratio with respect to $c$ and obtain
	\[ \frac{2}{1 - a \left(\frac{4}{1 - \frac{1}{c}} - 2\right)} - \frac{8a}{\left(1 - a \left(\frac{4}{1 - \frac{1}{c}} - 2\right)\right)^{2} \left(1 - \frac{1}{c}\right)^{2} c}\]
	which becomes $0$ when $c= \frac{2 \sqrt{2\alpha^{2} + \alpha} + 2\alpha + 1}{1-2\alpha}$, and we achieve a competitive ratio of
	\[\frac{2 (2 \alpha + \sqrt{\alpha (1 + 2 \alpha)})(1 + 2 \alpha + 2 \sqrt{\alpha (1 + 2 \alpha)})}{(1 - 2 \alpha)^2 \sqrt{\alpha (1 + 2 \alpha)}}\]
	as we expected.
\end{proof}

Observe that the results obtained for Algorithm~\ref{alg:submult} also apply for Algorithm~\ref{alg:sub} if we take the non-strict definition of the competitive ratio for any additive constant $\beta\ge \alpha$. This works as the only difference between the two algorithms is line~\ref{line:lowdensity} from Algorithm~\ref{alg:sub}, that rejects items of density at most $\alpha$. Those items can be packed into a knapsack to obtain a total value of at most $\alpha$, and thus can be offset by any chosen constant $\beta\ge \alpha$.

\section{Conclusions}
The general knapsack with reservation costs allows us to model situations where bookings can be canceled for some fee, in a more flexible setting than that of the online simple knapsack. We have presented upper and lower bounds for reservation costs depending on the size and value of the items. In the case of reservation costs depending on size, the bounds are tight, but some gaps remain in the case of reservations depending on item value.

\bibliography{knapsack}

\end{document}